\documentclass[aip,reprint,nofootinbib]{revtex4-1}
\usepackage{graphicx}
\usepackage{bm}

\begin{document}
\title{A Symplectic Method to Generate Multivariate Normal Distributions}
\date{\today}
\author{C. Baumgarten}
\affiliation{Paul Scherrer Institute, Switzerland}
\email{christian.baumgarten@psi.ch}

\def\begeq{\begin{equation}}
\def\endeq{\end{equation}}
\def\begary{\begeq\begin{array}}
\def\endary{\end{array}\endeq}
\def\bmtx{\left(\begin{array}}
\def\emtx{\end{array}\right)}
\def\eps{\varepsilon}
\def\d{\partial}
\def\y{\gamma}
\def\e{\eta}
\def\w{\omega}
\def\W{\Omega}
\def\s{\sigma}
\def\ket#1{\left|\,#1\,\right>}
\def\bra#1{\left<\,#1\,\right|}
\def\bracket#1#2{\left<\,#1\,\vert\,#2\,\right>}
\def\erw#1{\left<\,#1\,\right>}

\def\Exp#1{\exp\left(#1\right)}
\def\Log#1{\ln\left(#1\right)}
\def\Sinh#1{\sinh\left(#1\right)}
\def\Sin#1{\sin\left(#1\right)}
\def\Tanh#1{\tanh\left(#1\right)}
\def\Tan#1{\tan\left(#1\right)}
\def\Cos#1{\cos\left(#1\right)}
\def\Cosh#1{\cosh\left(#1\right)}

\begin{abstract}
The AMAS group at the Paul Scherrer Institute developed an object oriented library
for high performance simulation of high intensity ion beam transport with space charge~\cite{HPC1,HPC2}.
Such particle-in-cell (PIC) simulations require a method to generate multivariate particle distributions 
as starting conditions.

In a preceeding publications it has been shown that the generators of symplectic transformations
in two dimensions are a subset of the real Dirac matrices (RDMs) and that few symplectic transformations 
are required to transform a quadratic Hamiltonian into diagonal form~\cite{rdm_paper,geo_paper}.

Here we argue that the use of RDMs is well suited for the generation of
multivariate normal distributions with arbitrary covariances. A direct and simple
argument supporting this claim is that this is the ``natural'' way how such distributions
are formed. The transport of charged particle beams may serve as an example:
An uncorrelated gaussian distribution of particles starting at some initial position
of the accelerator is subject to linear deformations when passing through various
beamline elements. These deformations can be described by symplectic transformations.

Hence, if it is possible to derive the symplectic transformations that bring up these covariances,
it is also possible to produce arbitrary multivariate normal distributions without Cholesky decomposition. 
The method allows the use of arbitrary uncoupled distributions. The functional form of the coupled
multivariate distributions however depends in the general case on the type of the used random
number generator. Only gaussian generators always yield gaussian multivariate distributions.
\end{abstract}

%Hamiltonian mechanics, 45.20.Jj
%Oscillators coupled, 05.45.Xt
%Beam optics  41.85.-p
%Lorentz transformation 03.30.+p
%Statistics, 02.50.-r
\maketitle

%%%%%%%%%%%%%%%%%%%%%%%%%%%%%%%%%%%%%%%%%%%%%%%%%%%%%%%%%%%%%%%%%%%%%%%%%%%%%%%%
\section{Introduction}
%%%%%%%%%%%%%%%%%%%%%%%%%%%%%%%%%%%%%%%%%%%%%%%%%%%%%%%%%%%%%%%%%%%%%%%%%%%%%%%%

In Ref.~\cite{geo_paper} the author presented a so-called ``decoupling'' method
that is based on the systematic use the real Dirac matrices (RDMs) in coupled 
linear optics. The RDMs are constructed from four pairwise anti-commuting basic 
matrices with the ``metric tensor'' $g_{\mu\nu}=\mathrm{Diag}(-1,1,1,1)$,
formally written as:
\begeq
\y_\mu\,\y_\nu+\y_\nu\,\y_\mu=2\,g_{\mu\nu}\,.
\endeq
The remaining $12$ RDMs are constructed as products of the basic matrices
as described in the appendix.

The use of the RDMs enables to derive a straightforward method to transform
transport matrices, force matrices (``symplices'') {\it and} $\sigma$-matrices in 
such a way that the transformed variables are independent, i.e. decoupled.

The reverse is required to generate multivariate normal distributions:
A transformation that transforms linear independent distributions of
variables in such a way that a given covariance matrix is generated.
The idea therefore is the following: Generate a set of independent normally
distributed variables with given variances and apply the inverse of the
decoupling transformation derived from the desired covariance matrix.
This will couple the ``independent'' variables in exactly the desired
way. The presented scheme assumes an even number of variables since it is
based on canonical pairs, i.e. position $q_i$ and momentum $p_i$ - but
it is always possible to ignore one of those variables.

Since the method is based on pairs of canonical variables, the decoupling
scheme always treats two pairs of variables at a time, resulting in the use
of $4\times 4$-matrices. If more than four random variables are required,
the decoupling can be used iteratively in analogy to the Jacobi diagonalization
scheme for symmetric matrices~\cite{geo_paper}.

\section{Coupled Linear Optics}

In this section we give a brief summary of the major concept. 
Given the following Hamiltonian function
\begeq
H={1\over 2}\,\psi^T\,{\bf A}\,\psi\,,
\label{eq_Hamiltonian}
\endeq
where ${\bf A}$ is a symmetric matrix and $\psi$ is a state-vector
or ``spinor'' of the form $\psi=(q_1,p_1,q_2,p_2)^T$. The state vector
hence contains two pairs of canonical variables.
The equations of motion (EQOM) then have the familiar form
\begary{rcl}
\dot q_i&=&{\d H\over \d p_i}\\
\dot p_i&=&-{\d H\over \d q_i}\,,
\label{eq_eqom_classical}
\endary
or in vector notation:
\begary{rcl}
\dot\psi&=&\y_0\,\nabla_\psi\,H\\
        &=&{\bf F}\,\psi\\
\label{eq_eqom_general}
\endary
where the {\it force matrix} ${\bf F}$ is given as  ${\bf F}=\y_0\,{\bf A}$.
The matrix $\y_0$ is the symplectic unit matrix (sometimes labeled ${\cal J}$ 
or ${\cal S}$) and is identified with the real Dirac matrix $\y_0$ (see appendix). 
We define the symmetric matrix of second moments $\sigma$ containing
the variances as diagonal and the covariances as off-diagonal elements. 
The matrix ${\bf S}$ is simply defined as the product of $\sigma$ with $\y_0$:
\begeq
{\bf S}=\sigma\,\y_0\,.
\label{eq_Sdef}
\endeq
Both matrices, ${\bf F}$ and ${\bf S}$, fulfill the following equation 
(using $\y_0^T=-\y_0$ and $\y_0^2=-{\bf 1}$):
\begeq
{\bf F}^T=\y_0\,{\bf F}\,\y_0\,.
\label{eq_symplex}
\endeq
Matrices that obey Eq.~\ref{eq_symplex} have been named {\it symplices}, but
they are also called ``infinitesimally symplectic'' or ``Hamiltonian'' matrices~\cite{Talman}. 
Symplices allow superposition, i.e. any sum of symplices is a symplex, but
only the product of anti-commuting symplices is a symplex~\cite{rdm_paper}.

Any real-valued $4\times 4$-matrix ${\bf M}$ can be written as a linear combination of
real Dirac matrices (RDM): 
\begeq
{\bf M}=\sum\limits_{k=0}^{15}\,m_k\,\y_k\,.
\endeq
The RDM-coefficients $m_k$ can be computed from the matrix ${\bf M}$ by:
\begeq
m_k={\mathrm{Tr}(\y_k^2)\over 32}\,\mathrm{Tr}({\bf M}\,\y_k+\y_k\,{\bf M})\,,
\label{eq_rdmc}
\endeq
where $\mathrm{Tr}({\bf X})$ is the trace of ${\bf X}$.

Hence the RDMs form a complete system of all real $4\times 4$-matrices, but only 
ten RDMs fulfill Eq.~\ref{eq_symplex} and are therefore symplices: 
The basic matrices $\y_0,\dots,\y_3$ and the six ``bi-vectors'', i.e. the six 
possible products of two basic matrices. The symplices are the generators
of symplectic transformations, i.e. the generators of the symplectic group.

As well-known, the Jacobi matrix of a canonical transformation is symplectic,
i.e. it fulfills the following equation~\cite{Arnold,Talman}:
\begeq
{\bf M}\,\y_0\,{\bf M}^T=\y_0\,.
\label{eq_Msymplectic}
\endeq
The EQOM have the general solution
\begeq
\psi(t)={\bf M}(t,t_0)\,\psi(t_0)\,,
\label{eq_transsolution}
\endeq
where ${\bf M}$ is a symplectic {\it transfer matrix} that is in case of constant 
forces given by
\begeq
{\bf M}(t,t_0)=\exp{\left({\bf F}\,(t-t_0)\right)}\,.
\label{eq_transfer_matrix}
\endeq
Given now an (initial) set of $N$ normally distributed uncorrelated random variables $\psi_i$, then
the $\sigma$-matrix of these variables is given by
\begeq
\sigma={1\over N}\,\sum\limits_{i=0}^{N-1}\,\psi_i\,\psi_i^T\equiv\langle\psi\,\psi^T\rangle\,,
\endeq  
where the superscript ``T'' indicates the transpose, then the distribution
at time $t$ is given by:
\begeq
\sigma_t={1\over N}\,\sum\limits_{i=0}^{N-1}\,{\bf M}\,\psi_i\,\psi_i^T\,{\bf M}^T={\bf M}\,\sigma_0\,{\bf M}^T\,.
\endeq
Hence with Eqn.~(\ref{eq_Sdef}) and (\ref{eq_Msymplectic}) one has:
\begary{rcl}
{\bf S}_t&=&-{\bf M}\,\sigma_0\,\y_0^2\,{\bf M}^T\,\y_0\\
         &=&{\bf M}\,{\bf S}_0\,{\bf M}^{-1}\,.
\endary
That is - the transformation of ${\bf S}$ is a similarity-transformation 
with a symplectic transformation matrix. The reverse transformation obviously is
\begeq
{\bf S}_0={\bf M}^{-1}\,{\bf S}_t\,{\bf M}\,.
\endeq
Now we refer to the structural identity of the matrix ${\bf S}$ with
the force matrix ${\bf F}$. Both are symplices and since a transformation
that decouples ${\bf F}$ has been shown to diagonalize the matrix ${\bf A}$
of the Hamiltonian~\cite{rdm_paper,geo_paper},
it is clear that the same method can be used to diagonalize $\sigma$.
The reverse of this transformation then generates the desired distribution
from an initially uncorrelated $\sigma$.

Instead of a Cholesky-decomposition we may therefore use a symplectic
similarity-transformation to generate the correlated distribution 
from an initially uncorrelated distribution. In the context of charged particle
optics, the algorithm delivers even more useful information: the transformation
matrix ${\bf M}^{-1}$ is the transport matrix that is required to
generate an uncorrelated beam. 

\section{Symplectic Transformations and the Algorithm}

The general form of a symplectic transformation matrix ${\bf R}_b$ is that of 
a matrix exponential of a symplex $\y_b$ multiplied by a parameter $\eps$
representing either the angle or the ``rapidity'':
\begary{rcl}
{\bf R}_b(\eps)&=&\exp{(\y_b\,{\eps\over 2})}={\bf 1}\,c+\y_b\,s\\
{\bf R}_b^{-1}(\eps)&=&\exp{(-\y_b\,{\eps\over 2})}={\bf 1}\,c-\y_b\,s\,,
\endary
where 
\begary{rcl}
c&=&\left\{
\begin{array}{lp{10mm}lcr}
\cos{(\eps/2)}&for&\y_b^2&=&-{\bf 1}\\
\cosh{(\eps/2)}&for&\y_b^2&=&{\bf 1}\\
\end{array}\right.\\
s&=&\left\{
\begin{array}{lp{10mm}lcr}
\sin{(\eps/2)}&for&\y_b^2&=&-{\bf 1}\\
\sinh{(\eps/2)}&for&\y_b^2&=&{\bf 1}\\
\end{array}\right.\\
\label{eq_sincos}
\endary
Transformations with $\y_b^2=-{\bf 1}$ are orthogonal transformations, i.e. {\it rotations}, 
while those with $\y_b^2={\bf 1}$ are {\it boosts}.

The matrix ${\bf S}$ then is transformed according to:
\begeq
{\bf S}\to{\bf R}\,{\bf S}\,{\bf R}^{-1}\,.
\endeq
The decoupling requires a sequence of transformations, so that the 
RDM-coefficients of ${\bf S}$ have to be recomputed after each step.

Eqn.~\ref{eq_rdmc} may be used to compute the RDM-coefficients $s_k$
of the matrix ${\bf S}$
\begeq
{\bf S}=\sigma\,\y_0=\sum\limits_{i=0}^9\,s_k\,\y_k\,.
\endeq
Numerically it is faster to analyze directly the composition.
For the choice of RDMs used in Ref.~\cite{rdm_paper,geo_paper}
the RDM-coefficients of ${\bf S}$ as a function of $\sigma$ are given by: 
\begary{rcl}
s_0&=&(\sigma_{11}+\sigma_{22}+\sigma_{33}+\sigma_{44})/4\\
s_1&=&(-\sigma_{11}+\sigma_{22}+\sigma_{33}-\sigma_{44})/4\\
s_2&=&(\sigma_{13}-\sigma_{24})/2\\
s_3&=&(\sigma_{12}+\sigma_{34})/2\\
s_4&=&(\sigma_{12}-\sigma_{34})/2\\
s_5&=&-(\sigma_{14}+\sigma_{23})/2\\
s_6&=&(\sigma_{11}-\sigma_{22}+\sigma_{33}-\sigma_{44})/4\\
s_7&=&(\sigma_{13}+\sigma_{24})/2\\
s_8&=&(\sigma_{11}+\sigma_{22}-\sigma_{33}-\sigma_{44})/4\\
s_9&=&(\sigma_{14}-\sigma_{23})/2\\
\label{eq_rdm_coeffs}
\endary
Now we use the following abbreviation using the notation of $3$-dimensional
vector algebra:
\begary{rcl}
{\cal E}&=&s_0\\
\vec P&=&(s_1,s_2,s_3)^T\\
\vec E&=&(s_4,s_5,s_6)^T\\
\vec B&=&(s_7,s_8,s_9)^T\,,
\label{eq_emeq}
\endary
and furthermore:
\begary{rclp{5mm}rcl}
M_r&=&\vec E\,\vec B&&\vec r&\equiv&{\cal E}\,\vec P+\vec B\times \vec E \\
M_g&=&\vec B\,\vec P&&\vec g&\equiv&{\cal E}\,\vec E+\vec P\times \vec B \\
M_b&=&\vec E\,\vec P&&\vec b&\equiv&{\cal E}\,\vec B+\vec E\times \vec P \\
\label{eq_aux_vecs}
\endary
The decoupling is done by a sequence of maximal six symplectic transformations~\cite{geo_paper}.
A transformation with $\eps=0$ can be omitted. After each transformation,
the RDM-coefficients $s_k$ have to be updated and Eqns.~(\ref{eq_emeq}) and
~(\ref{eq_aux_vecs}) have to be re-evaluated:
\begin{enumerate}
\item ${\bf R}_0(\eps)$ with $\eps=\arctan{({M_g\over M_r})}$.
\item ${\bf R}_7(\eps)$ with $\eps=\arctan{({b_z\over b_y})}$.
\item ${\bf R}_9(\eps)$ with $\eps=-\arctan{({b_x\over b_y})}$.
\item ${\bf R}_2(\eps)$ with $\eps=\mathrm{artanh}{({M_r\over b_y})}$.
\item ${\bf R}_0(\eps)$ with $\eps={1\over 2}\,\arctan{({2\,M_b\over \vec E^2-\vec P^2})}$
\item ${\bf R}_8(\eps)$ with $\eps=-\arctan{({P_z\over P_x})}$.
\end{enumerate}

Given an initial covariance matrix $\sigma_0$, the sequence of computation 
therefore is:
\begin{enumerate}
\item Compute the RDM-coefficients $s_k$ according to Eqn.~(\ref{eq_rdm_coeffs}) and the quantities 
defined in Eqns.~(\ref{eq_emeq}) and ~(\ref{eq_aux_vecs}).
\item Compute the first (or next, resp.) transformation matrix ${\bf R}$.
\item Compute the product of the transformation matrices (and of the inverse) ${\bf M}_{n+1}={\bf R}_{n+1}\,{\bf R}_n$.
\item Apply the first (or next, resp.) transformation ${\bf S}_{n+1}={\bf R}\,{\bf S}_n\,{\bf R}^{-1}$.
\item Compute $\sigma_{n+1}=-{\bf S}_{n+1}\,\y_0$.
\item Continue with next transformation at step 1).
\end{enumerate}
The six iterations yield the desired diagonal matrix $\sigma_6$ and the matrices
${\bf M}_6$ and its inverse, so that
\begeq
{\bf S}_6={\bf M}_6\,{\bf S}_0\,{\bf M}_6^{-1}\,.
\endeq
or:
\begeq
\sigma_6={\bf M}_6\,\sigma_0\,{\bf M}_6^T\,.
\endeq
The diagonal elements of $\sigma_6$ are the variances of the
uncoupled gaussian distribution. Given $\psi_i$ is the i-th 
uncoupled random state vector, then ${\bf M}_6^{-1}\,\psi_i$ 
is the corresponding state vector with the multivariate 
normal distribution.

\section{Example}

Consider for instance the (arbitrary) matrix of second moments $\sigma_0$
{\small\begeq
\bmtx{cccccc}
 5.8269 & -0.0303 &  0.2292 &  0.0000 & -0.0960 &  1.4897 \\
-0.0303 &  0.8851 &  0.0000 & -0.0311 &  1.8053 & -0.0015 \\
 0.2292 &  0.0000 &  3.6058 & -0.0235 &  0.0000 &  0.0000 \\
 0.0000 & -0.0311 & -0.0235 &  0.6844 &  0.0000 &  0.0000 \\
-0.0960 &  1.8053 &  0.0000 &  0.0000 &  7.0607 & -0.0224 \\
 1.4897 & -0.0015 &  0.0000 &  0.0000 & -0.0224 &  0.7304 \\
\emtx
\endeq}
\begin{figure}
\includegraphics[width=7.5cm]{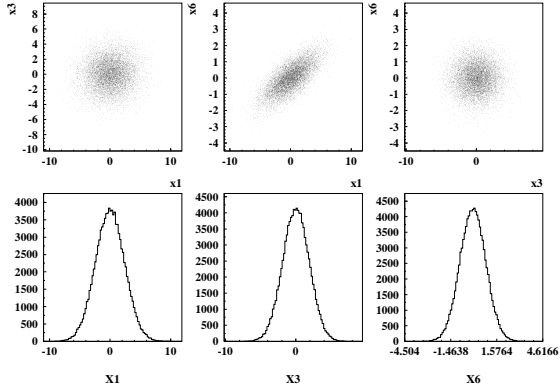}
\caption{
Top: Correlations between several variables of the multivariate normal distribution.
Bottom: The resulting probability distributions for the individual variables are
again gaussian.
\label{fig_gauss}
}
\end{figure}
The diagonal matrix $\sigma_6$ is computed to be
{\small\begeq
\bmtx{cccccc}
 1.9982 & 0 & 0 & 0 & 0 & 0 \\
 0 & 1.2984 & 0 & 0 & 0 & 0  \\
 0 & 0 &  1.1116 & 0 & 0 & 0 \\
 0 & 0 &  0 & 2.2124 & 0 & 0 \\
 0 & 0 &  0 & 0 & 1.4029 & 0 \\
 0 & 0 & 0 & 0 & 0 & 1.6637\\
\emtx
\endeq}
Now $10^5$ random vectors have been generated with a 
Gaussian random number generator of unit variance.
The vector elements have been scaled with corresponding
variances, given by the root of the diagonal elements
of $\sigma_6$ and then been multiplied (or transformed)
with ${\bf M}^{-1}$ given by
{\small\begeq
\bmtx{cccccc}
-0.1727 & -0.0330 &  0.0081 & -1.6049 & -0.0725 &  0.1893 \\
-0.0371 &  0.3392 &  0.7051 &  0.0093 &  0.3402 &  0.1034 \\
 0.9474 &  0.1485 &  0.0008 & -0.0613 & -0.3429 &  0.9838 \\
-0.0573 &  0.5025 & -0.0072 &  0.0001 & -0.4733 & -0.1464 \\
-0.1641 &  1.6387 &  0.3732 &  0.0202 &  1.4755 &  0.4320 \\
-0.3455 & -0.0555 &  0.0042 & -0.3515 & -0.1204 &  0.3416 \\
\emtx
\endeq}
Then the covariance matrix of the produced random vectors
was evaluated. The result is:
{\small\begeq
\bmtx{cccccc}
 5.7946 & -0.0247 &  0.2343 & -0.0060 & -0.1036 &  1.4825 \\
-0.0247 &  0.8917 &  0.0023 & -0.0306 &  1.8200 &  0.0034 \\
 0.2343 &  0.0023 &  3.5910 & -0.0299 & -0.0158 &  0.0033 \\
-0.0060 & -0.0306 & -0.0299 &  0.6849 & -0.0000 & -0.0012 \\
-0.1036 &  1.8200 & -0.0158 & -0.0000 &  7.0928 & -0.0148 \\
 1.4825 &  0.0034 &  0.0033 & -0.0012 & -0.0148 &  0.7294 \\
\emtx
\endeq}
Fig.~\ref{fig_gauss} shows some of the distributions as examples.

\begin{figure}
\includegraphics[width=7.5cm]{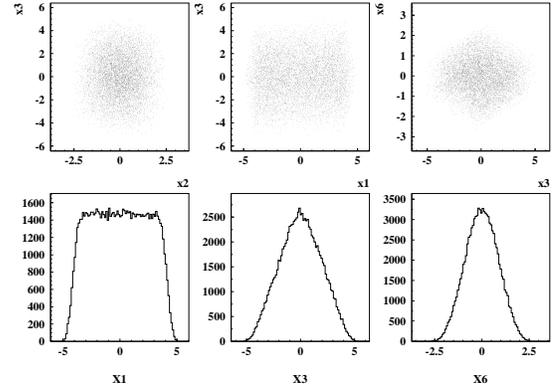}
\caption{
Top: Correlations between several variables of the multivariate flat distribution.
Bottom: The resulting probability distributions for the individual variables 
strongly depend on the correlations. The stronger the correlations, the more
gaussian the distribution will be (central limiting theorem).
\label{fig_flat}
}
\end{figure}

The same procedure can be done with any initial probability 
distribution and the algorithm will produce the desired 
second moments. But the functional form of the resulting 
distributions of the transformed variables will only be 
similar to the initial distribution in the Gaussian case.
Fig.~\ref{fig_flat} shows the results for the same covariance
matrix if the decoupled variables have a uniform probability 
distribution, but same variances. The covariance matrix
is correctly reproduced.

\section{Conclusion}

The method of symplectic decoupling of linearily coupled variables has been
applied to the problem of multivariate random distributions.
It has been shown that the use of sympleptic algebra has severe advantages:
The same methods can be applied to solve a variety of problems. The presented
algorithm is especially interesting for the generation of starting conditions
of particle tracking codes like - for example - OPAL~\cite{HPC1,HPC2}.

In cases where the decoupled process is known to have a non-Gaussian probability 
distribution and if the transport matrix ${\bf M}$ of a linear transport 
system is known, it should be possible to derive unknown para\-meters of the
initial distribution by comparison with the computed expected distribution. 
Fig.~\ref{fig_flat} shows that a flat distribution yields a clear ``signature''.

\begin{acknowledgments}
The software used for the computation has been written in ``C'' and been compiled 
with the GNU\textsuperscript{\copyright}-C++ compiler 3.4.6 on Scientific Linux.
The CERN library (PAW) was used to generate the figures.
\end{acknowledgments}

\begin{appendix}

\section{The $\y$-Matrices}
\label{sec_app1}
%%%%%%%%%%%%%%%%%%%%%%%%%%%%%%%%%%%%%%%%%%%%%%%%%%%%%%%%%%%%%%%%%%%%%%%%%%%%%%%%

The real Dirac matrices used throughout this paper are:
{\small
\begeq
\begin{array}{rclp{4mm}rcl}
\y_0&=&\bmtx{cccc}
   0 &   1  &  0 &   0\\
  -1 &   0  &  0 &   0\\
   0 &   0  &  0 &   1\\
   0 &   0  & -1 &   0\\
\emtx&&
 \y_1&=&\bmtx{cccc}
   0 &  -1  &  0 &   0\\
  -1 &   0  &  0 &   0\\
   0 &   0  &  0 &   1\\
   0 &   0  &  1 &   0\\
\emtx\\
 \y_2&=&\bmtx{cccc}
   0 &   0  &  0 &   1\\
   0 &   0  &  1 &   0\\
   0 &   1  &  0 &   0\\
   1 &   0  &  0 &   0\\
\emtx&&
 \y_3&=&\bmtx{cccc}
  -1 &   0  &  0 &   0\\
   0 &   1  &  0 &   0\\
   0 &   0  & -1 &   0\\
   0 &   0  &  0 &   1\\
\emtx\\
\y_{14}&=&\y_0\,\y_1\,\y_2\,\y_3;&&\y_{15}&=&{\bf 1}\\
\y_4&=&\y_0\,\y_1;&&\y_7&=&\y_{14}\,\y_0\,\y_1=\y_2\,\y_3\\
\y_5&=&\y_0\,\y_2;&&\y_8&=&\y_{14}\,\y_0\,\y_2=\y_3\,\y_1\\
\y_6&=&\y_0\,\y_3;&&\y_9&=&\y_{14}\,\y_0\,\y_3=\y_1\,\y_2\\
\y_{10}&=&\y_{14}\,\y_0=\y_1\,\y_2\,\y_3&&\y_{11}&=&\y_{14}\,\y_1=\y_0\,\y_2\,\y_3\\
\y_{12}&=&\y_{14}\,\y_2=\y_0\,\y_3\,\y_1&&\y_{13}&=&\y_{14}\,\y_3=\y_0\,\y_1\,\y_2\\
\end{array}
\endeq
}

\end{appendix}

\end{document}